\pdfoutput=1
\documentclass{aip-cp}

\usepackage[numbers]{natbib}
\usepackage{rotating}
\usepackage{graphicx}

\begin{document}

\title{Order, Chaos and (Quasi-) Dynamical Symmetries across 1st-order Quantum Phase Transitions in Nuclei}

\author[aff1]{M. Macek\corref{cor1}}
\author[aff2]{P. Cejnar}
\author[aff2]{P. Str\'ansk\'y}
\author[aff3]{J. Dobe{\v s}}
\author[aff4]{A. Leviatan}

\affil[aff1]{The Czech Academy of Sciences, Institute of Scientific Instruments, Brno, Czech Republic}
\affil[aff2]{Institute of Particle and Nuclear Physics, Charles University, Prague, Czech Republic}
\affil[aff3]{The Czech Academy of Sciences, Nuclear Physics Institute, {\v R}e{\v z}, Czech Republic}
\affil[aff4]{Racah Institute of Physics, The Hebrew University, Jerusalem 91904, Israel}
\corresp[cor1]{Corresponding author: michal.macek@isibrno.cz}

\maketitle

\begin{abstract}
First order quantum phase transition  (QPT) between spherical and axially deformed nuclei shows coexisting, but well-separated regions of regular and chaotic dynamics. We employ a Hamiltonian of the Arima-Iachello Interacting Boson Model (IBM) with an arbitrarily high potential barrier separating the phases. Classical and quantum analyses reveal markedly distinct behavior of the two phases: Deformed phase is completely regular, while the spherical phase shows highly chaotic dynamics, similar to the H\'enon-Heiles system. 
Rotational bands with quasi-SU(3) characteristics built upon the regular vibrational spectrum of beta- and gamma-vibrations are observed in the deformed phase up to very high excitation energies.
\end{abstract}
\vskip 10pt
Professor Iachello (whom we all know as Franco) has much enjoyed and has been virtuous in identifying aesthetic patterns, symmetries, in complicated phenomena of Nature. We know him as striving for elegant and unified understanding by formulating as simple models as possible (but not simpler). Perhaps curiously, Franco's models provide chances to study also the ``exact opposite'': mechanisms of symmetry breaking, (phase) transitions between different symmetries and even onset of chaos. Again curiously, Franco has been greatly supportive of people pursuing to study this fascinating ``exact opposite'' and has been one of the pioneers in it himself. 

The Interacting Boson Model (IBM)~\cite{ref:AIbook} combines three basic dynamical symmetries (DS) of nuclear collective motion: the U(5) DS of the spherical vibrator, the SU(3) DS of the axially-deformed rotor, and the SO(6) DS of the axially unstable rotor. In a generic case, the model allows mixing all three of these incompatible symmetries, providing possibilities to study highly non-trivial dynamics, yet with some symmetry-related ``handles'' to understand it (at least partially). In this short contribution, we would like to point out some of the features identified over the years when approaching the ``exact opposite'' of symmetry. We will show that at the first order quantum phase transition (QPT) between the U(5) and SU(3) dynamical symmetries~\cite{ref:Scholten}, a coexistence of completely regular and completely disordered dynamics is present in a single system, and even more strikingly in the same energy ranges: coexisting, yet clearly separated. Further, we will show that the regularity is not connected to an exact DS, but a quasi-dynamical symmetry (QDS), involving coherent linear combinations of irreducible representations of a DS~\cite{ref:Rowe}. In particular, this happens also throughout the phase coexistence region between the spinodal and anti-spinodal points~\cite{ref:LevMac}: The regular dynamics corresponds to the quasi SU(3) symmetry in the deformed phase, while the chaos---in its onset following the H\'enon-Heiles scenario~\cite{ref:HenonHeiles}---corresponds to the spherical phase~\cite{ref:MacLev}. 

An IBM Hamiltonian best suited to capture 1st order QPT behavior is the following pair:
\begin{eqnarray}\label{eq:Hs}
\hat{H}_1(\rho)/\bar{h}_2 &=& 2(1\!-\! \rho^2\beta_0^{2})\hat{n}_d(\hat{n}_d \!-\! 1) +\beta_0^2 R^{\dag}_2(\rho) \cdot\tilde{R}_2(\rho) ~,\qquad \label{eq:H1} \\
\hat{H}_2(\xi)/ \bar{h}_2 &=& \xi P^{\dag}_0(\beta_0) P_0(\beta_0) + P^{\dag}_2(\beta_0) \cdot \tilde{P}_2(\beta_0) ~,\label{eq:H2}
\end{eqnarray} 
where $\hat{n}_d = d^\dag\cdot\tilde{d}$ is the $d$-boson number operator, and the monopole and quadrupole pairing operators are $P^{\dag}_0 = d^{\dag}\cdot d^{\dag} - \beta_{0}^{2} (s^{\dag})^2$ and $P^{\dag}_{2\mu}(\beta_0) = \sqrt{2}\beta_0 s^\dag d^\dag_\mu + \sqrt{7}(d^\dag d^\dag)^{(2)}_\mu$ and $R^{\dag}_{2\mu}(\rho) = \sqrt{2} s^\dag d^\dag_\mu + \rho\sqrt{7}(d^\dag d^\dag)^{(2)}_\mu$. The coefficients $\rho, \xi$ are control parameters, while the overall scale is set here to $\bar{h}_2 = 2/N(N-1)$. $\hat{H}_1(\rho)$ is relevant for the spherical side of the QPT, while $\hat{H}_2(\xi)$ for the deformed side. The two Hamiltonians (\ref{eq:Hs}) coincide at the critical point $H_1(\rho=\beta_0^{-1}) = H_2(\xi = 0)$. Primary virtue of this apparently complicated formulation is that the last parameter, $\beta_0$, allows to directly adjust (i) the height of the phase-separating potential barrier at the critical point of the QPT, and (ii) the position (i.e. amount of deformation) of the deformed potential minimum~\cite{ref:MacLev}. Besides, this 
Hamiltonian possesses a peculiar symmetry property---the partial dynamical symmetry (PDS)~\cite{ref:PDS}: $H_1(\rho)$ has U(5) DS for $\rho = 0$ and U(5) PDS for any other $\rho \neq 0$, while $H_2(\xi)$ has SU(3) DS for $\xi = 1$ and $\beta_0 = \sqrt{2}$ and SU(3) PDS for $\beta_0 = \sqrt{2}$ and any values of $\xi\neq 1$; details in~\cite{ref:MacLev}. 

\begin{figure}[t]
  \centerline{\includegraphics[width=\linewidth]{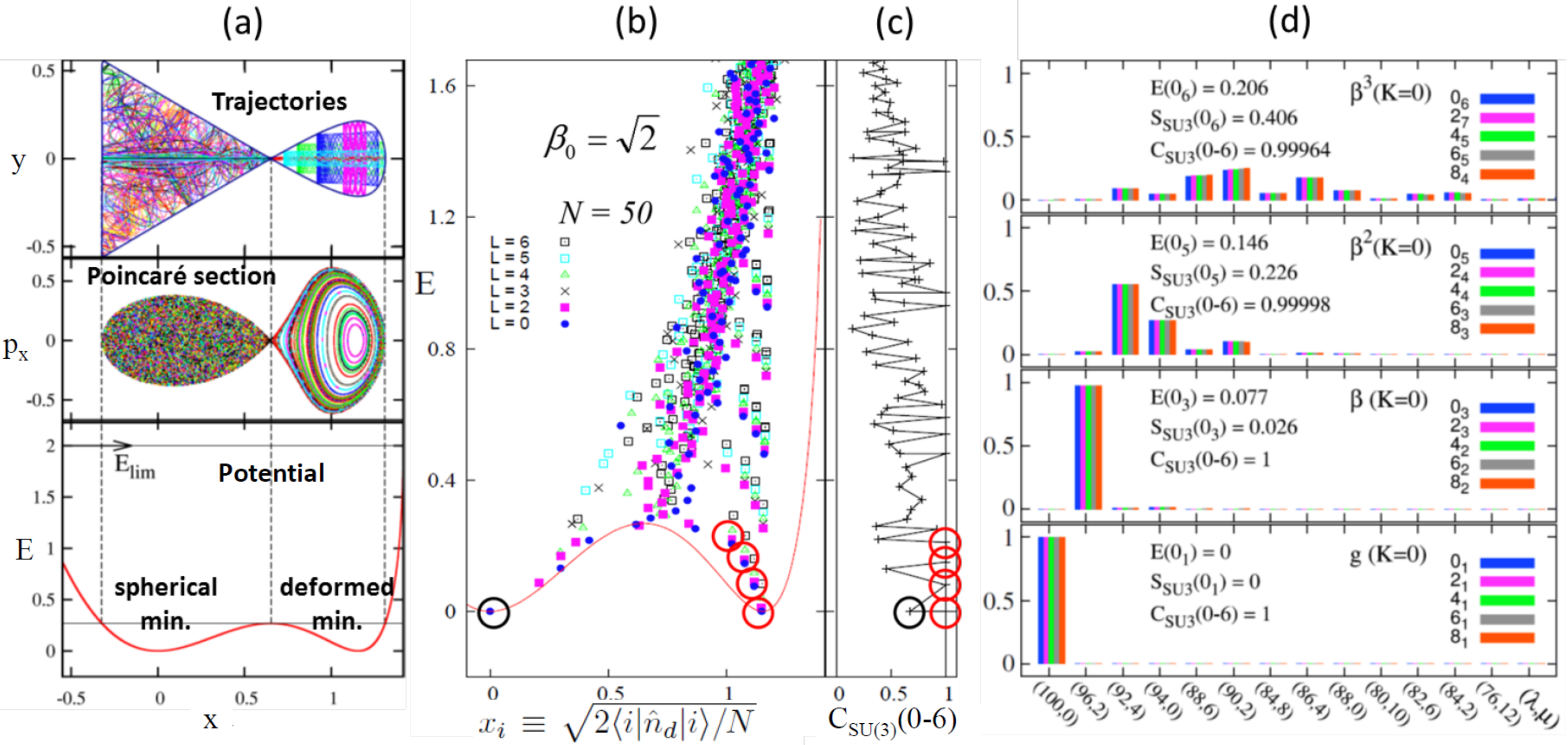}}
  \caption{Classical and quantum IBM dynamics at the critical point $\xi=0$ for $\beta_0=\sqrt{2}$. Panel (a): Classical trajectories, the Poincar\'e section and the potential $V$. Panel (b): Expectation values for $L = 0,2,3,4$ eigenstates $|i\rangle$ of the quantity $x_i =\sqrt{\langle i|\hat{n}_d |i\rangle/N}$ for total number of $N=50$ bosons embedded in the corresponding classical potential (red line). Panel (c): Correlation coefficient $\mathrm{C_{SU3}}(0-6)$ for angular momentum $L=0,2,4,6$ multiplets (see text) corresponding to states in panel (b). Panel (d): Decomposition of selected $K=0$ bands [g, $\beta$, $\beta^2$ and $\beta^3$, cf. red circles in panels (b,c)] in the SU(3) basis labeled by quantum numbers $(\lambda,\mu)$. Energy and SU(3)-basis Shannon entropy $\mathrm{S_{SU(3)}}(0_i)$ is indicated for each $L=0$ state, and the correlation coefficient $\mathrm{C_{SU(3)}}(0-6)$ for each $L=0,2,4,6$ multiplet. $\mathrm{S_{SU(3)}}(0_1)=0$ for the deformed ground band reflects the PDS, while the other multiplets show QDS with $\mathrm{S_{SU(3)}}(0_i)>0$.}
\end{figure}

Figure 1 snapshots the classical (panel a) and quantum (panels b-d) dynamics generated by $H_1(\rho = \beta_0^{-1}) = H_2(\xi=0)$, i.e. directly at the critical point of the QPT, for $\beta_0 = \sqrt{2}$; the latter allowing for the partial SU(3) DS. The classical dynamics is generated by Hamiltonians obtained as expectation values of (\ref{eq:Hs}) in Glauber coherent states~\cite{ref:AW,ref:ArcMoloch,ref:MacLev}. Two degenerate (spherical and deformed) minima of the classical potential $V(x,y=0)$ at the critical point are seen in panel (a,bottom). The dynamics related to both of them at energy corresponding to top of the phase-separating barrier (a saddle point of $V(x,y)$) is shown in the two panels above it: Trajectories evolving around the deformed minimum are regular, forming a set of (deformed) circles in the corresponding Poincar\'e section, while the trajectories around the spherical minimum are chaotic and fill the Poincar\'e section ergodically (Apart from periodic orbits, which form a measure zero set in the phase space.). 
Panel (b) shows an indicator of quantum chaos---a Peres lattice~\cite{ref:Peres} related to the quantity $x_i =\sqrt{\langle i|\hat{n}_d |i\rangle/N}$, which allows to associate the individual eigenstates  $|i\rangle$  with the classical potential $V(x,y=0)$. The lattices for eigenstates with angular momentum $L=0,2,3,4,5,6$ form regular patterns above the deformed minimum and (approximately) overlie each other, which reflects the presence of rotational bands with angular momentum projection on the symmetry axis $K=0,2,4,...$ here. Panel (c) reveals correlations between the $L=0,2,4,6$ eigenstates decomposed in the SU(3) DS basis by plotting a coefficient $\mathrm{C_{SU(3)}}(0-6)$, see~\cite{ref:Pearson}, as a function of energy of the $L=0$ ``bandheads'' (cf. panel b): The $K=0$ rotational bands identify by $\mathrm{C_{SU(3)}}(0-6)\approx 1$, while $L=0$ states not linked to any rotational structure have $\mathrm{C_{SU(3)}}(0-6)\ll 1$ (e.g. the spherical ground state at $E=0$). Panels (d) shows examples of some $K=0$ band members: in line with the correlation coefficient $\mathrm{C_{SU(3)}}(0-6)\approx 1$, their $(\lambda,\mu)$ distributions in the SU(3) basis are highly coherent, but apart from the ground band (which displays a SU(3) PDS), they do not fit within a single SU(3) irrep, expressing the SU(3) QDS~\cite{ref:Rowe}.     

\begin{figure}[ht]
  \centerline{\includegraphics[width=\linewidth]{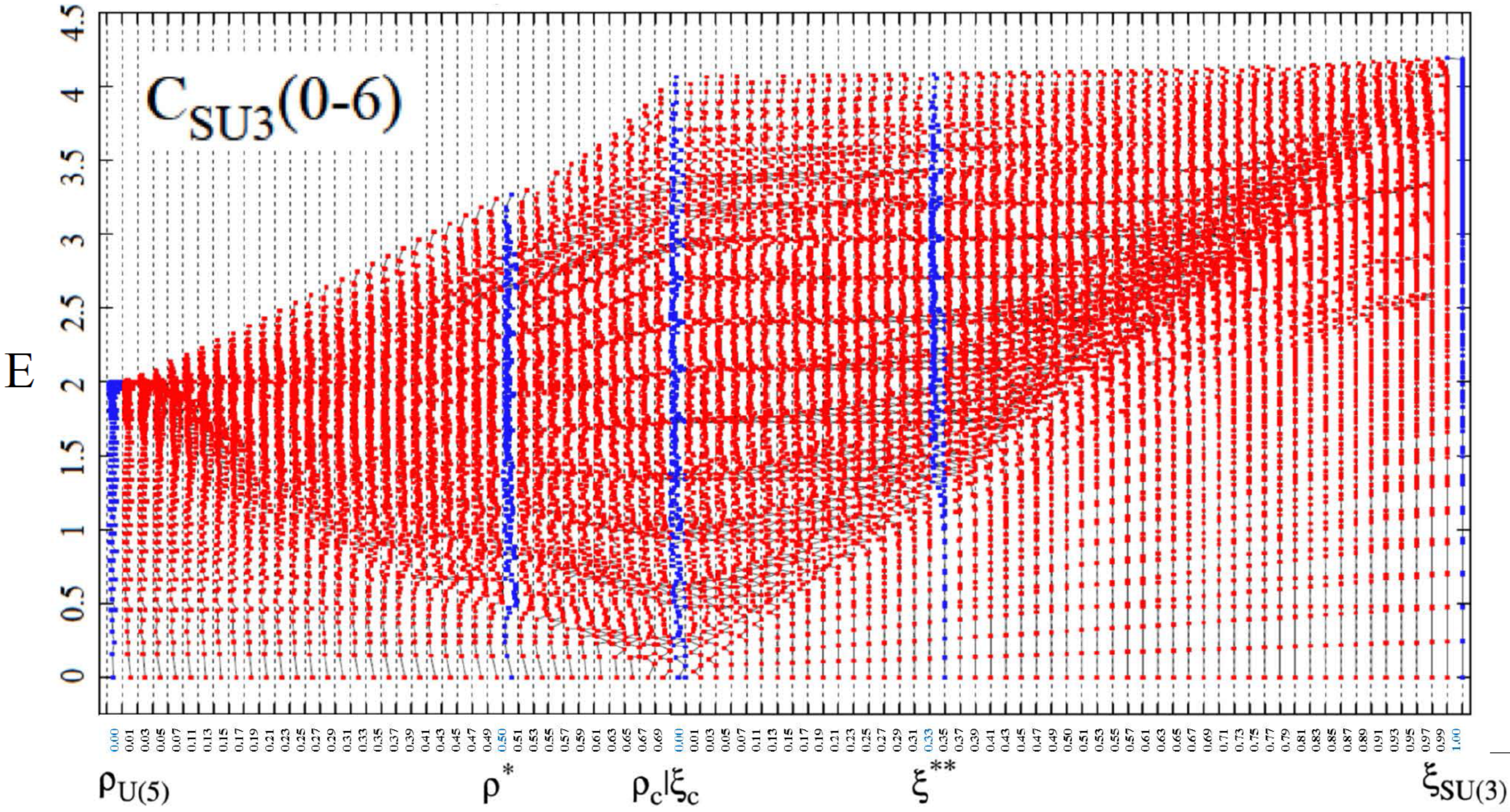}}
  \caption{Correlation coefficients $\mathrm{C_{SU3}}(0-6)$ as in Fig. 1 (c), however calculated at $86$ points between the U(5) DS ($\rho=0$) and the SU(3) DS ($\xi=1$) limits for $N=50$ bosons. Vertical dashed lines which embrace each control parameter, correspond to the values $\mathrm{C_{SU3}}(0-6)=0$ (left) and $\mathrm{C_{SU3}}(0-6)=1$ (right). Dependencies at DS limits, as well as the spinodal ($\rho^*$), critical ($\rho_c|\xi_c$) and anti-spinodal ($\xi^{**}$) points are highlighted by blue color. The critical point dependence is the same as in Fig. 1 (c).}
\end{figure}

In Fig. 2, we extensively display the evolution of SU(3) quasi dynamical symmetry across the 1st order QPT. The correlation coefficient $\mathrm{C_{SU(3)}}(0-6)$ shown in Fig. 1 (c) at the critical point is shown here in 86 points between the U(5) DS ($\rho=0$) and the SU(3) DS ($\xi=1$) limits. A major region where SU(3) QDS dominates the spectrum is seen on the deformed side of the QPT $\xi\geq 0$ at energies roughly below $E = 4\xi$. Interestingly, and consistent with the discussion above, there are extensive regions with SU(3) QDS elsewhere: Notice the multiple ``triangular features'' with SU(3) QDS values of $\mathrm{C_{SU3}}(0-6)\approx 1$ seen especially in the phase coexistence region (between $\rho^*$ and $\xi^{**}$) up to high energies above $E=2$. The ``triangular features'' are connected with finite-$N$ precursors of excited state quantum phase transition~\cite{ref:esqptII}. The fact that SU(3) QDS is so prolific and found even at very high excitation energy (c.f. the Alhassid-Whelan arc of regularity~\cite{ref:AW,ref:ArcMoloch}) may suggest that the related adiabatic separation of rotations and vibrations is due to an underlying regular motion of vibrational dynamics~\cite{ref:Adia}. 

The authors thank Franco Iachello for lasting inspiration, support and friendship.


\nocite{*}
\bibliographystyle{aipnum-cp}%
{}

\end{document}